\documentclass[preprint,12pt]{elsarticle}
\usepackage{bm}
\usepackage{amsmath}
\usepackage{amssymb}
\usepackage{xcolor,soul}
\usepackage{siunitx}
\usepackage{hyperref}
\usepackage{booktabs}
\usepackage{upgreek}
\usepackage{siunitx}
\newcolumntype{P}[1]{>{\centering\arraybackslash}p{#1}}

\journal{arXiv}

\begin{document}

\begin{frontmatter}

\title{Non-linear jog-dragging effect on the mobility law of edge dislocations in face-centered cubic nickel}

\author[1,2]{Wu-Rong Jian\fnref{equal}}

\author[2,3]{Yifan Wang\fnref{equal}}

\author[2]{Wei Cai\corref{corr-author}}
\ead{caiwei@stanford.edu}

\address[1]{Department of Engineering Mechanics, South China University of Technology, Guangzhou Guangdong, 510640, P. R. China}
\address[2]{Department of Mechanical Engineering, Stanford University, Stanford CA, 94305, USA}
\address[3]{Department of Materials Science and Engineering, Stanford University, Stanford CA, 94305, USA}

\fntext[equal]{Both authors contributed equally to the paper}
\cortext[corr-author]{Corresponding author}

\date{\today}


%
\begin{abstract}
%
Dislocation jogs have strong effects on dislocation motion that governs the strain-hardening behavior of crystalline solids,
but how to properly account for their effect in mesoscale models remains poorly understood.
We develop a mobility model for jogged edge dislocations in FCC nickel,
based on systematic molecular dynamics (MD) simulations across a range of jog configurations, stresses, and temperatures.
At low stresses, jogged edge dislocations exhibit non-linear, thermally activated dragging and a higher Peierls barrier compared to straight dislocations. 
Surprisingly, stress-velocity curves for a given jog configuration across varying temperatures intersect at an invariant point ($\tau_{\rm c}$, $v_{\rm c}$),
where $\tau_{\rm c}$ delineates thermally-activated and phonon-drag regimes and is close to the Peierls stress ($\tau_{\rm p}$).
Motivated by this observation, we propose a simple three-section expression for jogged dislocation mobility, featuring minimal and physically interpretable parameters.
This mobility law offers a realistic description of jog effects for dislocation dynamics (DD) simulations, improving their physical fidelity for crystal plasticity predictions.
\end{abstract}

\begin{keyword}
Jog effects \sep edge dislocation \sep dislocation mobility \sep face-centered cubic \sep molecular dynamics
\end{keyword}

\end{frontmatter}

\section{Introduction}

Jogs, appearing as offsets on a dislocation line by one or more interplanar spacings out of its original glide plane, are fundamental to dislocation dynamics.
They significantly influence the plasticity behaviors of crystalline materials, such as strain hardening and climb~\cite{hirth1982theory,hull2011introduction}.
During strain hardening, the intersection of dislocations on different slip systems results in the formation of jogs, which act as strong pinning points and significantly impede glide and contribute to the hardening.
It has long been recognized that dislocation jogs have strong dragging effects to dislocation motion from the theoretical analysis~\cite{messerschmidt1970pssb, messerschmidt1971pssb, messerschmidt2010dislocation}
On the other hand, dislocation jogs can also form during thermally activated dislocation climb~\cite{abu-odeh2020acta} or through point-defect-dislocation interactions~\cite{rodney2000prb, wang2025scripta},
enabling the dislocation to overcome obstacles in the original glide plane through the absorption or emission of vacancies~\cite{jian2022acta}.
Despite studies on jog formation mechanisms due to dislocation interactions during multiplication\cite{hirth1982theory,kondo2025scripta},
a quantitative description remains unclear for jog dynamics and its contribution to the dislocation mobility (velocity).

Accurate prediction of jogged dislocation mobility is crucial for robust discrete dislocation dynamics (DDD) simulations, which aim to model dislocation evolution and macroscopic stress-strain response~\cite{bulatov2006computer, bertin2020armr}.
Current DDD simulations, especially concerning dislocation self-organization during plastic deformation~\cite{arsenlis2007msmse}, 
frequently oversimplify jog behavior,
by either completely neglecting all jog effects~\cite{bertin2020armr}, or modeling jogs as completely sessile segments with zero mobility~\cite{li2023acta} in certain cases.
This omission of jog effects on dislocation mobility is hypothesized to be the cause of underestimated flow stress in DDD simulations compared to experimental observations~\cite{akhondzadeh2023acta}.
However, it is impossible to include such effects in DDD modeling without a systematic and quantitative study of the jog effects.
Therefore, developing a quantitative model of jog effects on dislocation mobility is crucial for DDD simulations to accurately describe the physics of dislocation motion and correctly predict the mechanical responses of crystalline materials.

As a key parameter to DDD simulations, dislocation mobility law is defined as dislocation velocity as a function of applied stress.
This quantity is often obtained through molecular dynamics (MD) simulations,
which track dislocation positions over time and extract dislocation velocities under different applied stresses.
MD simulations have successfully established mobility laws for straight dislocations in various metals, such as in aluminum~\cite{cho2017ijp,dang2019acta},
nickel~\cite{Olmsted2005msmse}, iron~\cite{queyreau2011prb}, molybdenum~\cite{chang_molecular_2002} and magnesium~\cite{groh2009msmse}.
However, our understanding of jogged dislocation mobility at the atomic scale remains limited.
While most existing MD investigations focus on straight dislocations, an exception was a study conducted in nickel~\cite{rodney2000prb}, which predicted the velocity of an edge dislocation with a pair of unit jogs as a function of applied stress at 100~K.
This research showed that a unit jog pair introduces significant additional friction, making the jogged dislocation immobile until a critical stress is reached. 

Nonetheless, the detailed effects of jog spacing (or density) along the dislocation line and jog height on the dislocation mobility law are still largely unexplored. 
Furthermore, given that jogs lead to a threshold-stress behavior, the conservative (glide) motion of jogged dislocation is likely thermally activated below this threshold, suggesting a pronounced temperature dependence.
This thermally activated glide may also be related to thermally activated dislocation climb~\cite{abu-odeh2020acta}.
To accurately reproduce jog behaviors and their impact on flow stress and strain hardening in DDD simulations, a quantitative dislocation mobility law is indispensable.
Constructing such a model necessitates a comprehensive understanding of how jog spacing, height, and temperature variations collectively influence jogged dislocation mobility.

In this work, we use molecular dynamics (MD) simulations to quantitatively explore how jog characteristics, specifically the height and the spacing along the dislocation line, influence edge dislocation mobility in face-centered cubic (FCC) nickel across temperatures ranging from $\qtyrange{100}{600}{K}$. 
We compare the glide mobility of jogged edge dislocations with that of their straight counterparts.
For a straight dislocation, the velocity-stress relationship exhibits two regimes: a low-stress regime where the velocity is linear with stress, and a high-stress regime where the velocity saturates due to relativistic effects.
In contrast, jogged dislocations display a non-linear thermally activated regime at low stresses, followed by a linear phonon-drag dominated regime and a relativistic saturation regime at high stresses.
We find that the drag effect is predominantly due to the constricted jog within the jog pair, while the extended jog is highly mobile and has minimal impact on the dislocation mobility.
Importantly, we find that all velocity-stress curves for a given jog geometry, measured at different temperatures, consistently pass through an invariant point, characterized by a critical velocity $v_{\rm c}$ at a critical stress $\tau_{\rm c}$.
This critical stress is very close to the Peierls stress $\tau_{\rm p}$, defined as the minimum stress required for dislocation motion at zero temperature.
Further analysis confirms that at stress below $\tau_{\rm c}$, jogged dislocation motion is indeed a stress-driven, thermally activated process, with an intrinsic energy barrier that increases with higher jog density and larger jog height. 
Building on this mechanistic understanding, we developed a quantitative, analytic model that describes the relationship between jogged dislocation velocity and applied stress across both the thermally activated and phonon-drag regimes. 
This mobility law encodes the essential physics of jogged dislocation motion and is suitable for use in larger-scale DDD simulations for crystal plasticity.

The remainder of this paper is organized as follows.
Section~\ref{sec:methodology} describes the MD simulation setup and analysis methods.
Section~\ref{sec:results} presents the main findings.
Section~\ref{sec:summary_of_results} summarizes the functional form of the mobility law for jogged edge dislocations, followed by a detailed description of the MD results in Section~\ref{sec:MD_results}.
Section~\ref{sec:fitted_parameters} describes how the parameters of the mobility law are fitted to the MD data.
Section~\ref{sec:discussion} provides further discussions on the physics of jog dragging.
Section~\ref{sec:discussion_geometry} examines the structures of the jogs and identifies the effect of jog dragging on the shape of the dislocation.
%
Section~\ref{sec:Peierls_stress} demonstrates the energy barrier of the jogged dislocation motion, showing a clear stress-dependent thermally activated kinetics.
Section~\ref{sec:conclusions} summarizes our conclusions and discusses the prospect of applying this mobility law in DDD simulations.

\section{Methods}
\label{sec:methodology}

\begin{figure}[!ht]
    \centering
    \includegraphics[width=0.85\linewidth]{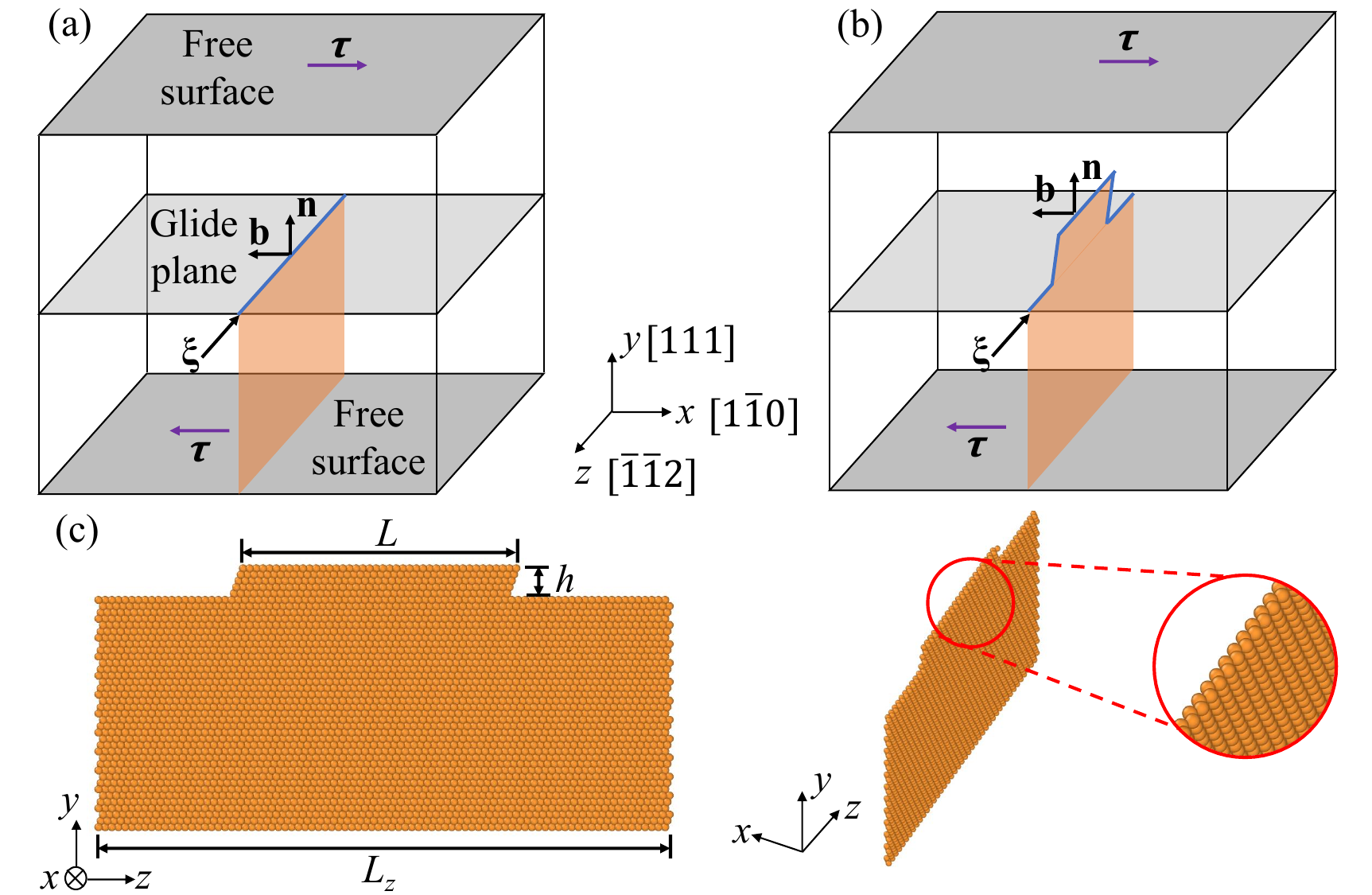}
    \caption{Schematic of MD simulation cell containing (a) a straight edge dislocation without jogs and (b) a jogged edge dislocation with the number of jog pair $N=1$.  The dislocation line is shown in blue. $\bm{\upxi}$, $\mathbf{b}$ and $\mathbf{n}$ denote dislocation line vector, Burgers vector and normal vector of the glide plane, respectively. The orange regions in (a) and (b) represent the half-plane of atoms that are removed to generate a dislocation. (c) The removed atomic layer viewed along two different angles to create the jogged edge dislocation configuration.}
    \label{fig:model_schematic}
\end{figure}


MD simulations are conducted using the LAMMPS simulation package~\cite{thompson2022cpc} employing the embedded-atom method (EAM) interatomic potential developed by Angelo \emph{et al.}~\cite{angelo1995msmse}.
The simulation cell has the edge vectors in the $x$, $y$, $z$ directions equal to the crystallographic vectors $96[1\bar{1}0]$, $24[111]$, and $84[\bar{1}\bar{1}2]$, respectively.  This results in a simulation box of approximately $24\times15\times\qty{36}{nm}^{3}$.
A single straight edge dislocation is introduced at the center of $x$-$y$ plane by removing a half-plane of atoms, as shown in \autoref{fig:model_schematic}(a).
The dislocation line is aligned with the $z$-axis, and its Burgers vector along the $x$-axes, resulting in a slip plane normal to the $y$-axis.
Periodic boundary conditions are applied along the $x$- and $z$- directions, and two free surfaces are introduced normal to the $y$-direction.

To introduce jog pairs on the straight dislocation, addition lines of atoms slightly above the dislocation slip plane (i.e. in the climb direction) are removed, as shown in \autoref{fig:model_schematic}(b).
The jog configuration is specified by two parameters: jog spacing $L$ and jog height $h$.
$L$ is calculated as $L = L_z/(2N)$, where $L_z$ is the simulation cell length in the $z$-direction, and $N$ is the number of introduced jog pairs.
Consequently, the jog density is thus defined as $\rho = 2N/L_z$.
The parameter $h$ represents the number of $(111)$ planes separating the glide planes on the two sides of the jog, indicating the height of the jogs. 
Note that the jog line is tilted to lie on the $(11\bar{1})$ slip plane (see \autoref{fig:model_schematic}(c)).
For convenience, the straight dislocation configuration is simplify referred to as ``straight''.
Jogged dislocation configurations are named based on their $L$ and $h$ values (e.g., $L$18$h$1 and $L$9$h$5).
Here the $L$ values are in the unit of $\si{nm}$, and $h$ values are the number of $\frac{1}{3}[111]$ interplanar spacings.
All jog configurations created for this study are summarized in \autoref{tbl:jogged_dislocation}.
Each configuration is then relaxed to achieve a stable core structure with dissociated partial dislocations, as further discussed in Section~\ref{sec:discussion_geometry}.

\begin{table}[!htb] 
\centering
\small
\caption{Jog spacings, densities and heights for all jogged dislocation configurations built in our MD simulations. The units for jog spacing, density and height are $\si{nm}$, $10^{9}$ m$^{-1}$ and number of $\frac{1}{3}[111]$ interplanar spacings, respectively.} \label{tbl:jogged_dislocation}
\begin{tabular}{ P{3.2cm} P{1.2cm} P{1.2cm} P{1.2cm} P{1.2cm} P{1.2cm} P{1.2cm} }
\toprule
Type  & $L$18$h$1 & $L$9$h$1 & $L$9$h$3 & $L$9$h$5 & $L$6$h$1 & $L$3$h$1 \\
\midrule
Spacing $L~(\si{nm})$   & 18 & 9 & 9 & 9 & 6 & 3 \\
Density $\rho~(\si{1/nm})$  & 0.056 & 0.111 & 0.111 & 0.111 & 0.167 & 0.333 \\
Height $h$   & 1 & 1 &3 & 5 & 1 & 1 \\
\bottomrule 
\end{tabular}
\end{table}

The relaxed dislocation configuration is then heated  to target temperatures of $\qty{100}{K}$, $\qty{300}{K}$, and $\qty{600}{K}$.  The NVT ensemble is used and box size is iteratively adjusted for $\qty{300}{ps}$ to reach equilibrium at zero stress.
To drive dislocation motion, a shear stress $\tau$ with the $xy$-component is applied.  This is achieved by applying external forces in the $+x$ direction to atoms on the top surface (with $+y$ normal vector) and in the $-x$ direction to atoms on the bottom surface (with $-y$ normal vector), as sketched in \autoref{fig:model_schematic}(a)-(b). 
For each applied stress value $\tau$, the dislocation velocity is measured from the slope of the dislocation position-time curve (Section~\ref{sec:result_disp_time}), after the dislocation motion has reached a steady state.
This process allows us to establish a dislocation mobility law $v(\tau)$, specifying the dislocation velocity $v$ as a function of the applied stress $\tau$.

\section{Results}
\label{sec:results}

\subsection{Summary of results}
\label{sec:summary_of_results}

\begin{figure}[!ht]
    \centering
    \includegraphics[width=0.95\linewidth]{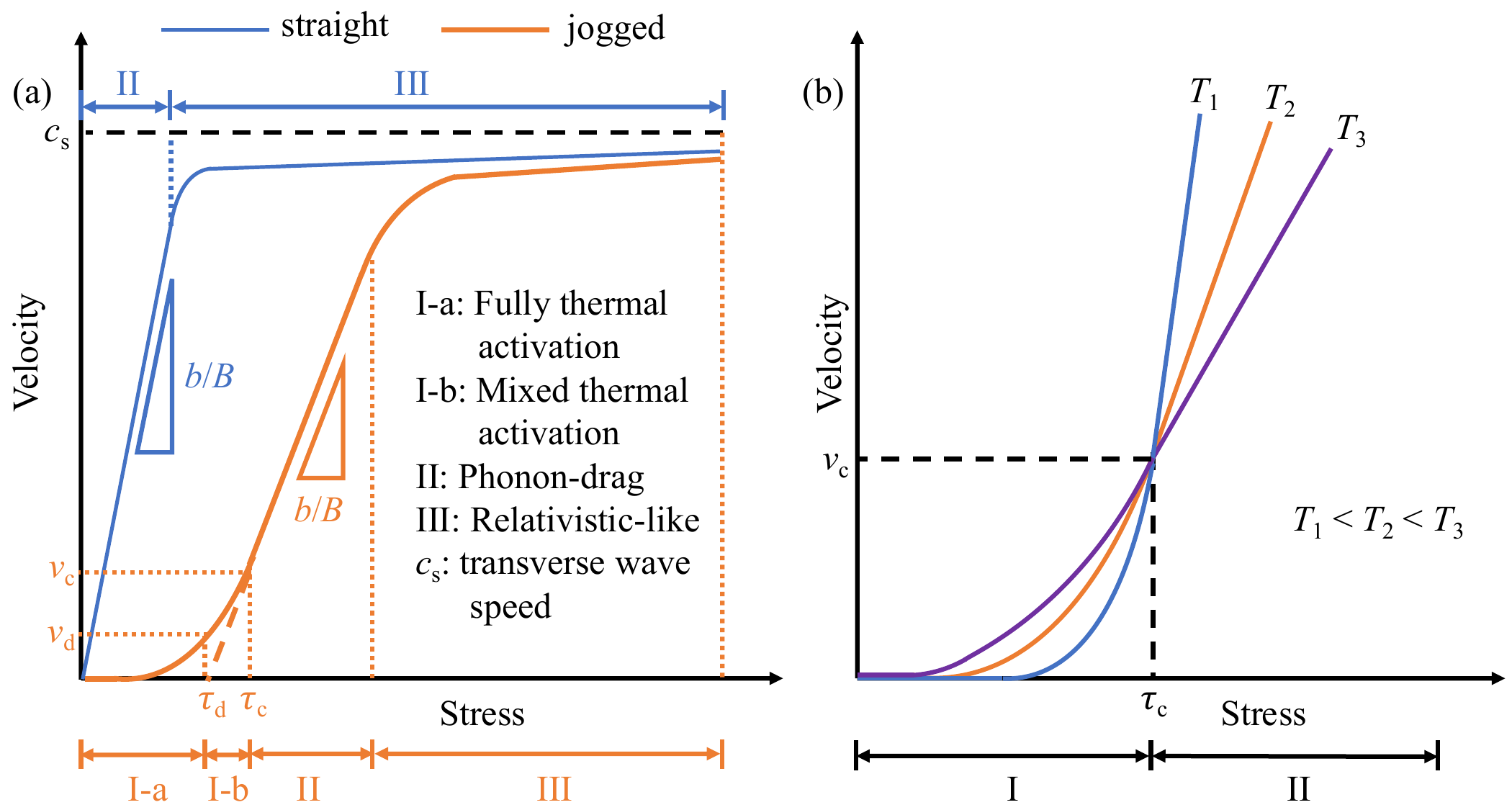}
    \caption{
    (a) Schematic for the velocity-stress relationships of straight and jogged edge dislocations, demonstrating two-regime and three-regime behaviors, respectively.
    $c_{\rm s}$ represents the transverse wave speed and the velocity limit due to relativistic effects.
    (b) Schematic for the velocity-stress curves of the jogged edge dislocation at different temperatures. 
    The velocity-stress curves appear to all pass through an invariant point, defined by a critical velocity $v_{\rm c}$ at a critical stress $\tau_{\rm c}$.
    }
    \label{fig:v_tau_schematic}
\end{figure}
\autoref{fig:v_tau_schematic} summarizes the mobility law of jogged edge dislocations derived from our MD simulations.
For comparison, \autoref{fig:v_tau_schematic}(a) also sketches the mobility law of straight edge dislocations (thin blue curve), 
which exhibit two regimes: a linear regime at lower stresses and a non-linear regime at higher stresses where the velocity saturates due to relativistic effects.
The presence of jogs significantly reduces the mobility of edge dislocation (thick orange curve), which displays a distinct three-regime behavior.
In Regime I, at the lowest stress range, jogged edge dislocation motion presents non-linear thermally activated motion.
In Regime II, within the medium stress range, the velocity increases linearly with stress due to the phonon-drag mechanism.
Finally, in Regime III, at the highest stress range, the velocity saturates towards the limiting velocity $c_{\rm s}$ due to relativistic effects.

The velocity-stress curves for jogged edge dislocations exhibit opposite temperature dependence in different stress regimes.
In Regime I (thermally activated regime), the dislocation velocity increases with increasing temperature,
while in Regime II (phonon-drag regime), the dislocation velocity decreases with increasing temperature.
Notably, for a given jog geometry (jog height $h$ and jog spacing $L$), all velocity-stress curves obtained at different temperatures consistently pass through an invariant point,
characterized by a critical velocity $v_{\rm c}$ at a critical stress $\tau_{\rm c}$, as illustrated in \autoref{fig:v_tau_schematic}(b).
The existence of this invariant point conveniently delineates the two regimes of dislocation motion,
and further signifies that the transition from one regime to another as stress increases is independent of temperature.
Such an invariant-point behavior has not been previously observed for other types of dislocation motion in metals with a transition from thermally activated to phonon-drag regimes,
such as screw or mixed dislocations in BCC metals~\cite{kang_singular_2012}.

The invariant point $(\tau_{\rm c}, v_{\rm c})$ can be conveniently used to parameterize the mobility law for the jogged edge dislocation.
For example, in Regime II, the mobility law is a linear function,
\begin{equation}
v(\tau, T) = v_{\rm c} + \frac{b}{B(T)} (\tau - \tau_{\rm c}) 
 \, , \quad \quad {\rm for} \ \tau \geq \tau_{\rm c}
    \label{eqn:phonon_drag_regime}
\end{equation}
where $B(T)$ is a temperature-dependent drag coefficient,
and can be further expressed as a linear function of temperature: $B(T) = B_0+B_1T$~\cite{po2016acta}.

To parameterize the mobility law in Regime I, we split it into two sub-regimes: I-a where thermal activation is the dominant mechanism, and I-b where both thermal activation and phonon-drag are expected to play a role.
The boundary between these two sub-regimes is defined as the intersection point ($\tau_{\rm d}$) of extrapolating the linear mobility law in Regime II with the horizontal axis, as shown in \autoref{fig:v_tau_schematic}(a).
In Regime I-a, the mobility law is written as the difference of two Arrhenius expressions of forward and backward jumps,
\begin{equation}
 v(\tau, T) = v_{\rm c}\left[
                \exp\left(-\frac{\Delta H_{\rm f}(\tau)}{k_{\rm B}T}\right)
              - \exp\left(-\frac{\Delta H_{\rm b}(\tau)}{k_{\rm B}T}\right) \right]
    \, , \quad \quad {\rm for} \ \tau < \tau_{\rm d}
    \label{eqn:thermally_activated_regime}
\end{equation}
where $\Delta H_{\rm f}$ and $\Delta H_{\rm b}$ are the activation enthalpies for forward and backward jumps, and are written in the Kocks form~\cite{mecking1981acta} below.
\begin{align}
    \Delta H_{\rm f}(\tau) = H_0^{\rm p} \left[ 1 -
            \left( \frac{\tau}{\tau_{\rm c}} \right)^p
        \right]^q \label{eq:DHf}\\
    \Delta H_{\rm b}(\tau) = H_0^{\rm p} \left[ 1 +
            \left( \frac{\tau}{\tau_{\rm c}} \right)^p
        \right]^q
        \label{eqn:DHf_DHb}
\end{align}
where $p$ and $q$ are fitting parameters, $H_0^{\rm p}$ is the Peierls barrier of the jogged dislocation at zero stress.
Finally, in Regime I-b, the mobility law is expressed as a cubic spline, to ensure a smooth transition from Regime I-a to Regime II, where both the velocity and its derivative with respect to stress are continuous,
\begin{align}
    v(\tau, T) = v_{\rm c} + Q_1\Delta\tau + Q_2\Delta\tau^2 + Q_3\Delta\tau^3
    \, , \quad \quad {\rm for} \ \tau_{\rm d} \leq \tau < \tau_{\rm c}
\end{align}
Here we define $\Delta\tau = \tau-\tau_c$, with the coefficients,
\begin{align}
    Q_1 &= B(T) \\
    Q_2 &= \frac{3(v_{\rm d} - v_{\rm c})}{(\tau_{\rm c}-\tau_{\rm d})^2} + \frac{s_{\rm d} + 2B(T)}{\tau_{\rm c}-\tau_{\rm d}} \\
    Q_3 &= \frac{2(v_{\rm d} - v_{\rm c})}{(\tau_{\rm c}-\tau_{\rm d})^3} + \frac{s_{\rm d} + B(T)}{(\tau_{\rm c}-\tau_{\rm d})^2}
    \label{eqn:Q_3}
\end{align}
where $v_{\rm d}=v(\tau_{\rm d})$ and $s_{\rm d} = (\partial v/\partial\tau)_{\tau=\tau_{\rm d}}$ are the velocity and its derivative values at the right boundary of Regime I-a, $\tau_{\rm d}$, which can be evaluated using \autoref{eqn:thermally_activated_regime}.

In this work, we focus on constructing a dislocation mobility law in Regimes I and II, as these are the most relevant stress regimes for DDD simulations of strain hardening, where the stress is often too low for relativistic effects to be significant.
For each jog configuration in \autoref{tbl:jogged_dislocation}, the seven parameters of the mobility law ($v_{\rm c}$, $\tau_{\rm c}$, $H_0^{\rm p}$, $p$, $q$, $B_0$, $B_1$) are fitted to the MD data, which are presented below.  Parameters $p$ and $q$ are kept constant across all jog configurations.

\subsection{MD results of jogged dislocation mobility}
\label{sec:MD_results}

During the MD simulation of dislocation motion at finite temperature $T$ and applied stress $\tau$, the perfect dislocation in nickel dissociates into partial dislocations.
We use the dislocation extraction analysis (DXA) in the OVITO~\cite{stukowski2012msmse} software to obtain the dislocation positions during the MD simulation.
To accurately determine the mobility law, we keep track of the dislocation displacement in the $x$-direction as a function of time, 
using the average $x$-position of the partial dislocation segments from DXA.

\subsubsection{Displacement-time curves}
\label{sec:result_disp_time} 

\begin{figure}[!ht]
    \centering
    \includegraphics[width=0.9\linewidth]{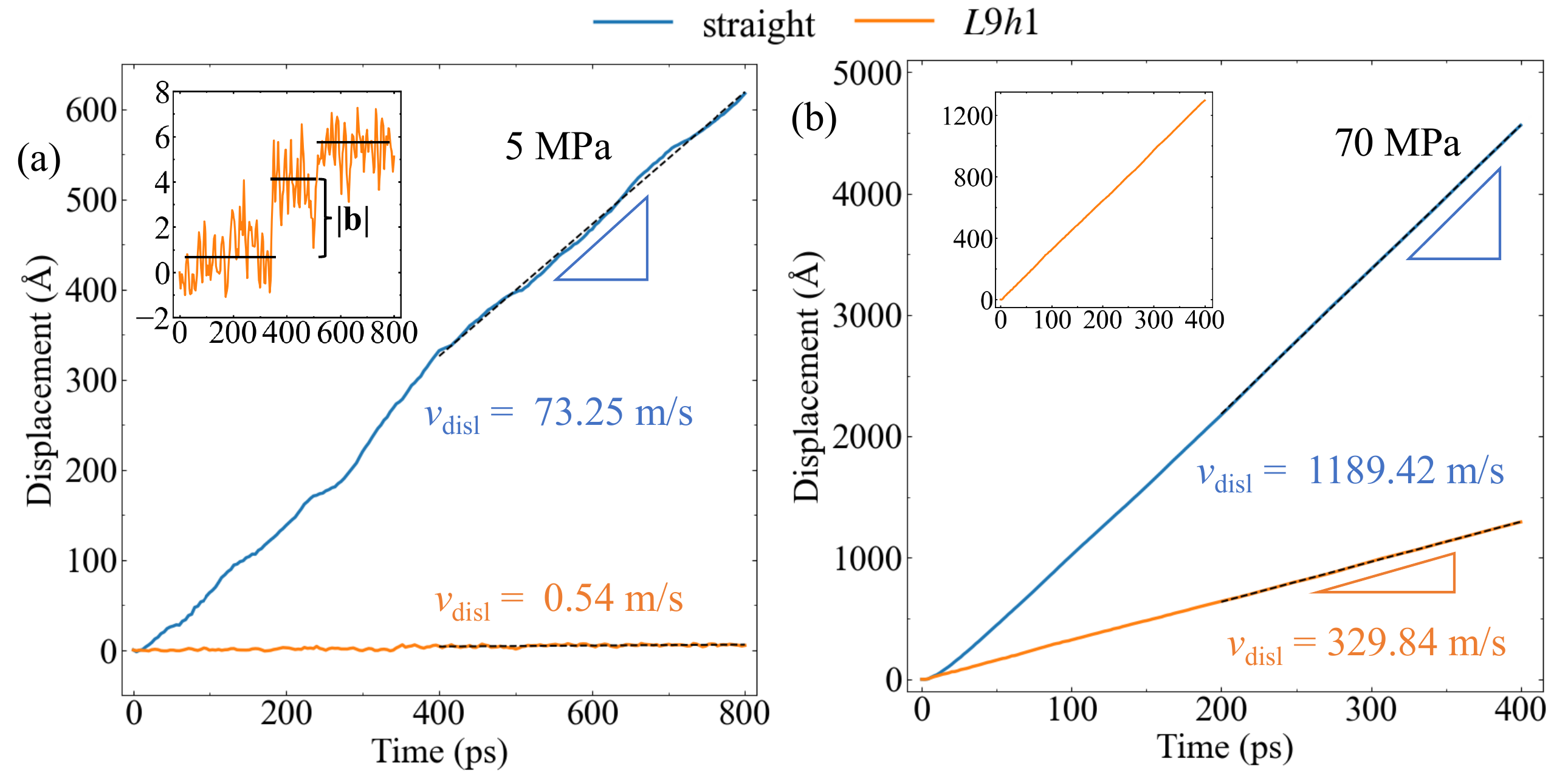}
    \caption{Dislocation displacement as a function of time for ``straight'' and ``$L9h1$'' dislocation configurations at 300 K, under shear stresses of (a) $\qty{5}{MPa}$ and (b) $\qty{70}{MPa}$, respectively. Here, ``straight'' and ``$L9h1$'' denote the straight and jogged dislocation configurations defined in Table~\ref{tbl:jogged_dislocation}.}
    \label{fig:D_t}
\end{figure}

\autoref{fig:D_t} shows the dislocation displacement in the glide direction as a function of time for both ``straight'' and ``$L9h1$'' configurations at $\qty{300}{K}$,
under applied stresses of $\qty{5}{MPa}$ and $\qty{70}{MPa}$, respectively.
The MD simulations of dislocation glide were long enough (up to $\qty{800}{ps}$~\cite{wang2025scripta}) to ensure that the steady-state motion has been achieved.
The velocity $v_{\rm disl}$ is then obtained by linear fitting the second half of the displacement-time curve.
A clear jog-dragging effect is evident: the jogged dislocation glides significantly slower than the straight edge dislocation. 
Interestingly, the ratio of the velocities between the straight and jogged dislocations is reduced as the applied stress increases,
indicating a reduced dragging effect at higher stresses.

The inset of \autoref{fig:D_t}(a) shows that at $\qty{5}{MPa}$ the jogged dislocation exhibits a step-wise, jerky motion, indicating that it is trapped in metastable states between discrete jumps.
This is a manifestation of a thermally activated mechanism, which we have labeled as Regime I.
At higher stress, such as \qty{70}{MPa} in \autoref{fig:D_t}(b), the jerky motion is replaced by smooth motion characteristic of the phonon-drag mechanism, which we have labeled as Regime II.

\subsubsection{Invariant point in velocity-stress curves}
\label{sec:invariant}

\begin{figure}[!ht]
    \centering
    \includegraphics[width=0.98\linewidth]{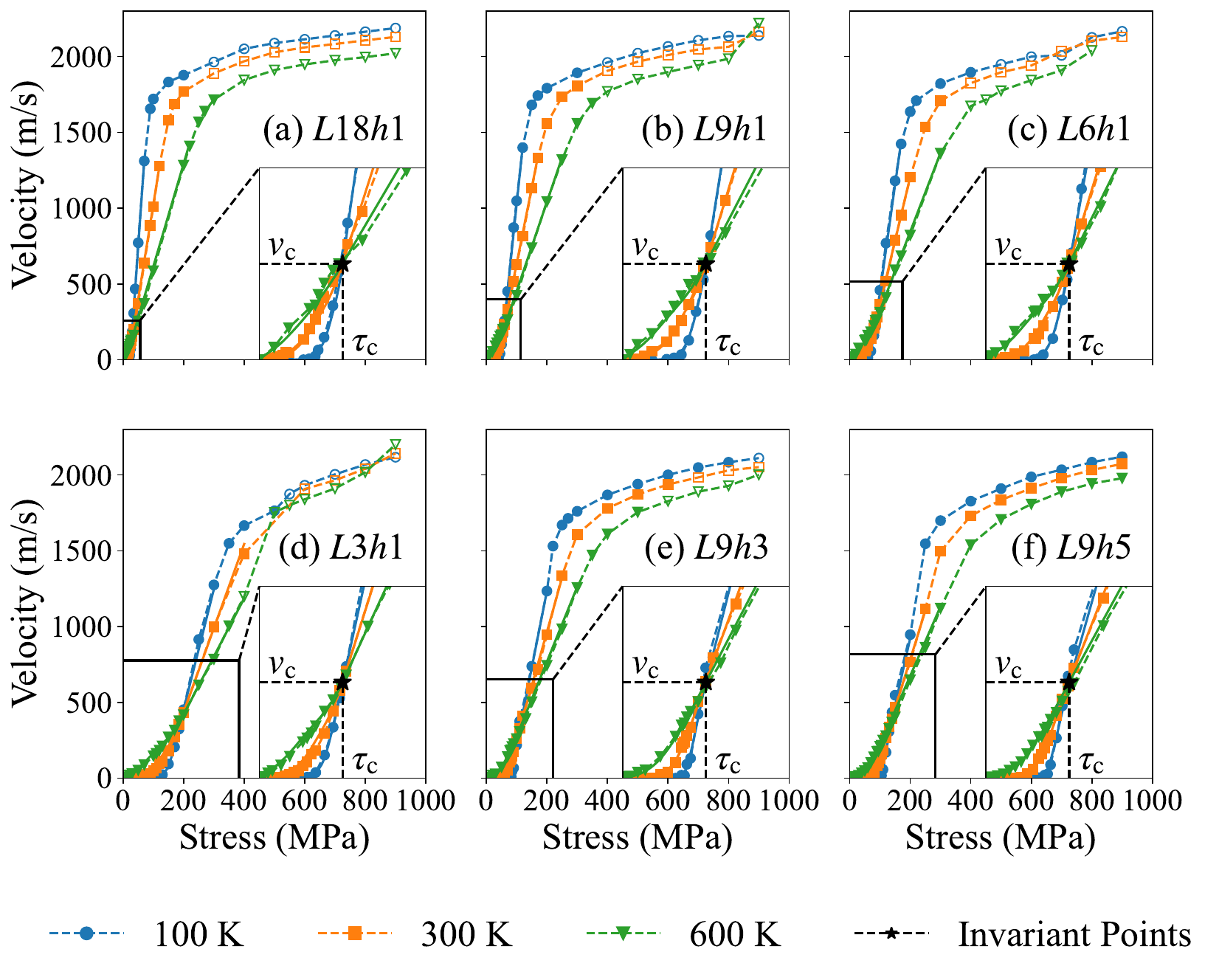}
    \caption{
    Temperature effect on dislocation velocity as a function of applied shear stress for all jogged edge dislocation configurations: (a) $L18h1$, (b) $L9h1$, (c) $L6h1$, (d) $L3h1$, (e) $L9h3$ and (f) $L9h5$.
    The solid lines indicate the fitted model discussed in Section~\ref{sec:summary_of_results}.
    The insets are the magnified curves at the low stresses,
    showing the invariant point of the mobility curves at different temperatures.
    The open symbols indicate the cases where vacancy emission is observed (see Section~\ref{sec:discussion_geometry}).
    }
    \label{fig:temperature}
\end{figure}

\autoref{fig:temperature} shows the dislocation velocities extracted from our MD simulations using the procedure described above, as a function of the applied shear stress for all six jogged dislocation configurations.
In each plot, dislocation velocity-stress curves at three temperatures: $\qty{100}{K}$, $\qty{300}{K}$ and $\qty{600}{K}$ are shown together.
Notably, for each jogged configuration, the velocity-stress curves at different temperatures all pass through the same point, $(v_{\rm c}, \tau_{\rm c})$, which we call the invariant point.
At stresses below the invariant point, increasing temperature leads to higher dislocation velocities, consistent with a thermally activated mechanism.
At stresses above the invariant point, increasing temperature leads to lower dislocation velocities, consistent with a phonon drag mechanism.
The overall shapes of the velocity-stress curves observed here provide the basis of the schematics shown in Figure~\ref{fig:v_tau_schematic}, including the definition of the three stages of dislocation motion.
Our mobility law focuses on modeling the dragging effect at the low-stress regimes relevant to strain hardening (Regime I \& II).
The invariant point serves as a convenient boundary separating Regime I and Regime II.

\subsection{Fitted mobility law parameters}
\label{sec:fitted_parameters}

\begin{table}[htb] 
\centering
\small
\caption{
Parameters for the dislocation mobility law, Eqs.~(\ref{eqn:phonon_drag_regime})-(\ref{eqn:Q_3}), fitted to MD simulations of different jog configurations. The parameters $p = 0.53$ and $q = 1.32$ are kept constant across all jog configurations. The temperature dependence of the drag coefficient in Regime II is expressed by $B(T)=B_0+B_1\,T$.  The Peierls stress $\tau_{\rm p}$ is also listed to compare with the fitted parameter $\tau_{\rm c}$.
} 
\label{tbl:Peierls_Vc}
\begin{tabular}{ P{3.5cm} P{1.3cm} P{1.3cm} P{1.3cm} P{1.3cm} P{1.3cm} P{1.3cm} }
\toprule
  & $L$18$h$1 & $L$9$h$1 & $L$9$h$3 & $L$9$h$5 & $L$6$h$1 & $L$3$h$1 \\
\midrule
$\tau_{\rm p}$ (MPa)  & 30 & 59 & 114 & 143 & 89 & 188 \\
$\tau_{\rm c}$ (MPa)  & 28.29 & 56.67 & 110.37 & 141.87 & 87.12 & 191.60 \\
$v_{\rm c}$ (m/s)   & 127.74 & 197.90 & 325.92 & 410.98 & 258.78 & 396.42 \\
$H_0^{\rm p}$ (eV) & 0.22 & 0.22 & 0.34 & 0.34 & 0.22 & 0.22 \\
$B_0$ ($\mathrm{\mu}$Pa$\cdot \mathrm{s}$)   & 3.14 & 7.16 & 19.38 & 22.97 & 11.37 & 24.82 \\
$B_1$ ($\mathrm{\mu}$Pa$\cdot\mathrm{s}\cdot\mathrm{K}^{-1}$)   & 0.0560 & 0.0587 & 0.0518 & 0.0525 & 0.0608 & 0.0675 \\
\bottomrule 
\end{tabular}
\end{table}

\begin{figure}[!ht]\centering\includegraphics[width=0.99\linewidth]{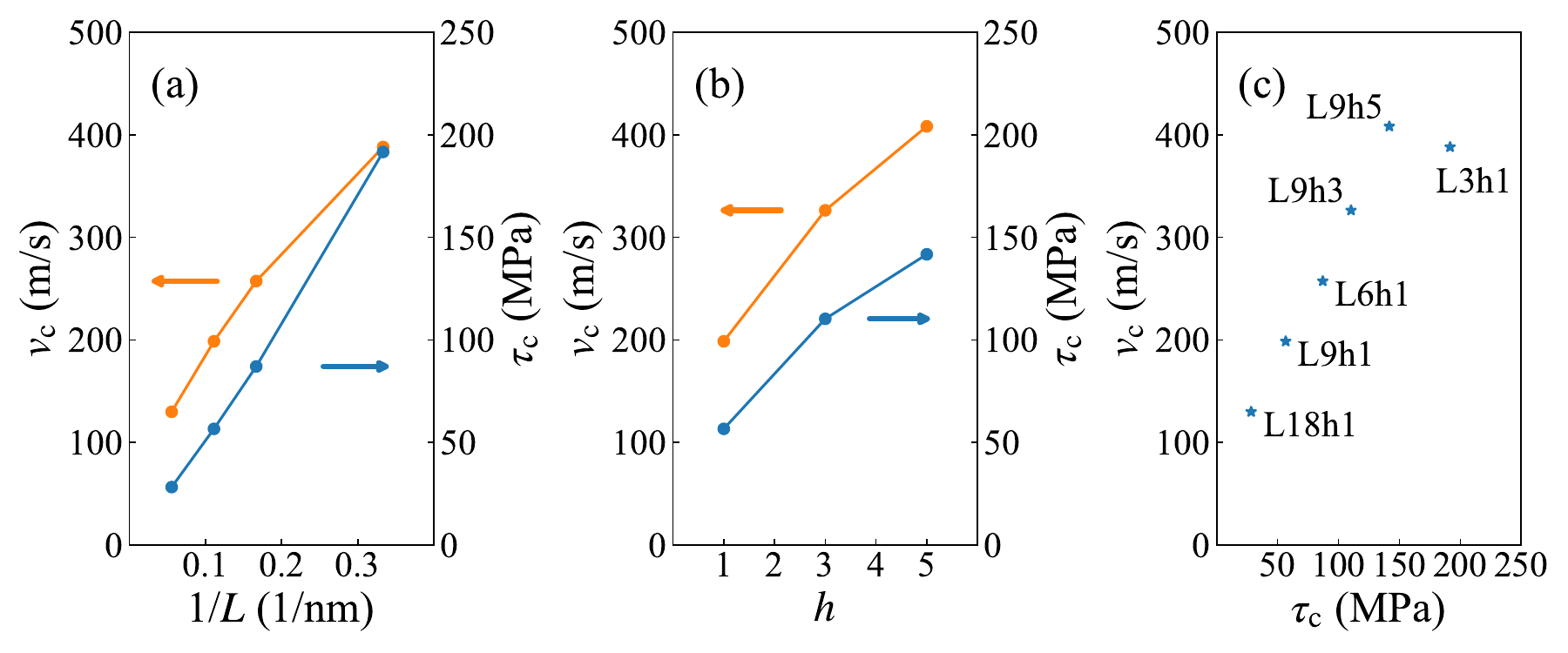}
    \caption{
    (a) Fitted $\tau_{\rm c}$ and $v_{\rm c}$ as functions of jog density $1/L$ for unit jog ($h = 1$);
    (b) Fitted $\tau_{\rm c}$ and $v_{\rm c}$ as functions of jog height $h$ for the same dislocation density (where $L = 9$ nm);
    (c) Fitted $\tau_{\rm c}$ and $v_{\rm c}$ for different jog configurations.
    }
    \label{fig:taup_Lh}
\end{figure}

The dislocation velocity as a function of stress and temperature extracted from MD simulations has been fitted to mobility expressions, Eqs.~(\ref{eqn:phonon_drag_regime})-(\ref{eqn:Q_3}), which cover the Regimes I (thermally activated) and II (phone drag).
The fitted parameters are listed in Table~\ref{tbl:Peierls_Vc}.
For each jogged dislocation configuration, we first determine the invariant point $(v_{\rm c}, \tau_{\rm c})$ where velocity-stress curves of all temperatures pass through.
We separately calculate the Peierls stress $\tau_{\rm p}$ using molecular statics for different jogged dislocation configurations, as the critical stress required to move the dislocation at zero temperature.
$\tau_{\rm p}$ is also listed in Table~\ref{tbl:Peierls_Vc} to show that it is close to the critical stress $\tau_{\rm c}$ of the invariant point.
\autoref{fig:taup_Lh} shows that (a) $\tau_{\rm c}$ is mostly linear with the jog density $1/L$, (b) $\tau_{\rm c}$ roughly linear with the jog height $h$, and (c) $\tau_c$ and $v_{\rm c}$ are more or less linear with each other among the jogged dislocation configurations considered here.

\begin{figure}[!ht]
    \centering
    \includegraphics[width=0.9\linewidth]{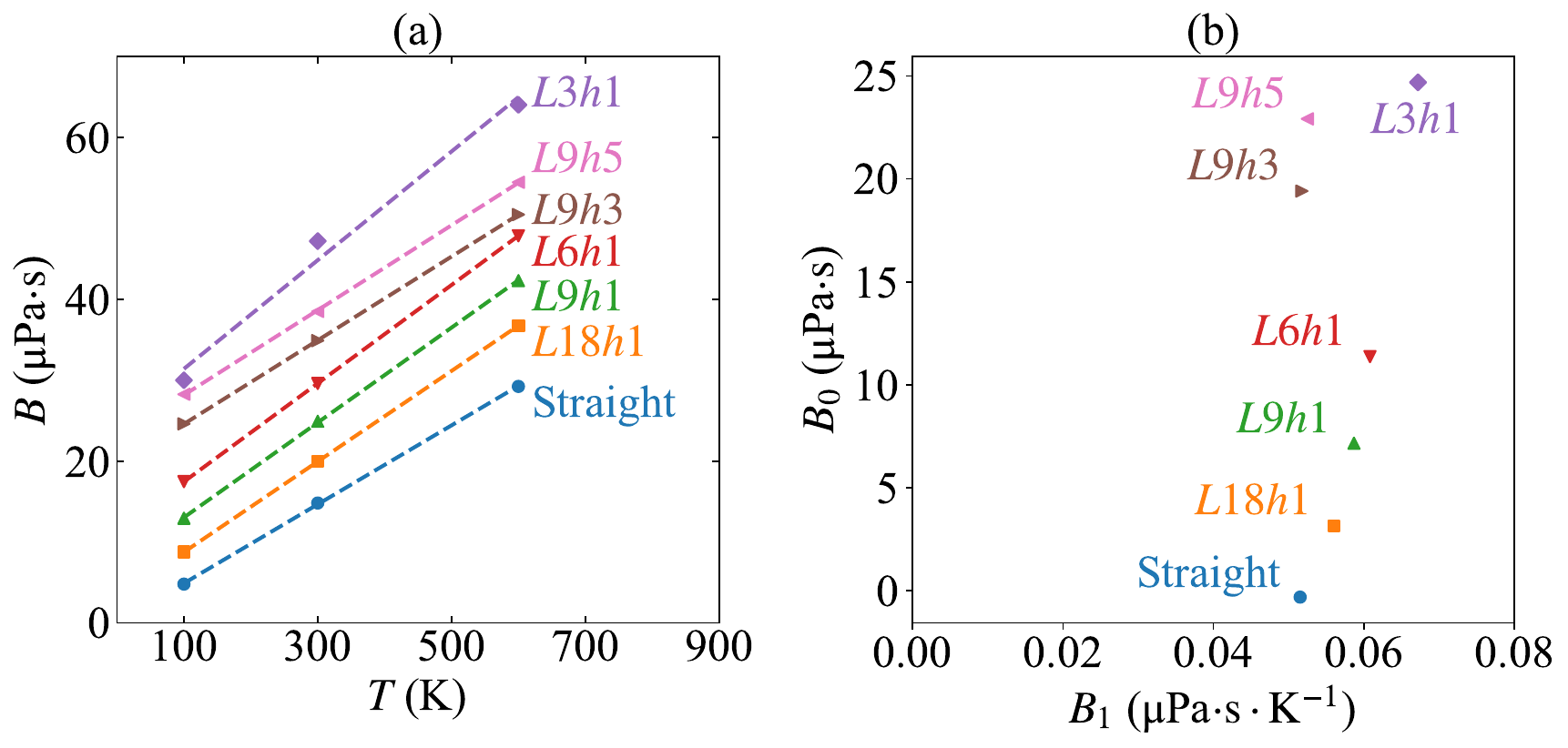}
    \caption{
        (a) Drag coefficient $B$ as a function of temperature $T$ for different dislocation configurations, fitted to $B(T) = B_0 + B_1\,T$ (dashed lines).
        (b) Coefficient $B_0$ and $B_1$ for different jog configurations ($L$ and $h$).
    }
    \label{fig:drag_coefs}
\end{figure}

We then determine the drag coefficient $B(T)$ in Regime II ($\tau > \tau_{\rm c})$ from the slope of the velocity-stress curves when they are linear (see solid lines in Figure~\ref{fig:temperature}).
\autoref{fig:drag_coefs} shows that $B(T)$ is well described by a linear function of temperature, $B(T) = B_0 + B_1\,T$.  The fitted coefficients $B_0$ and $B_1$ are listed in Table~\ref{tbl:Peierls_Vc}.
It is seen that while $B_0$ varies a lot with the jogged dislocation configuration.
The magnitude of $B_1$ stays mostly the same, but is smaller for the cases of the super-jog ($h>1$), corresponding to a weaker dependence of the drag coefficient on temperature (see \autoref{fig:drag_coefs}(b)).

Finally, we fit the velocity-stress curves in Regime I (thermally activated) using Eqs.~(\ref{eqn:thermally_activated_regime})-(\ref{eqn:DHf_DHb}).
Parameter $H_0^{\rm p}$ is the Peierls barrier, which is the energy barrier for dislocation motion at zero applied stress.
We calculated the $H_0^{\rm p}$ value of $\qty{0.22}{eV}$ for the $L18h1$ case with nudge-elastic band (NEB) method using the LAMMPS software package~\cite{thompson2022cpc}.
We kept the same $H_0^{\rm p}$ value for all unit jog ($h = 1$) cases because we assume the energy barrier for the jogged dislocation to move by one lattice distance at zero stress is independent of jog spacing.
%
On the other hand, the $H_0^{\rm p}$ value for the superjog cases ($h=3$ and $h=5$) are fitted to MD data and they are significantly higher than the value for unit jogs.
%
%
For simplicity, we use the same $H_0^{\rm p}$ value ($0.34$~eV) for the cases of $h=3$ and $h=5$.
This is justified by the observation that the dissociated jog structures look similar for the $h=3$ and $h=5$ cases.
The MD simulations also revealed that the dragging effect comes primarily from one side of the constricted node.  Hence, it is reasonable to assume that for large enough jog height $h$ the effect should be independent of $h$.
%
%
%
%
%
%
%
%
The exponents ($p$, $q$), sometimes called the Kocks' parameters, are kept constant for all jogged dislocation configurations in this work.

\section{Discussion}
\label{sec:discussion}

\subsection{Dislocation geometry of jog pairs}
\label{sec:discussion_geometry}

\begin{figure}[!ht]
    \centering
    \includegraphics[width=0.9\linewidth]{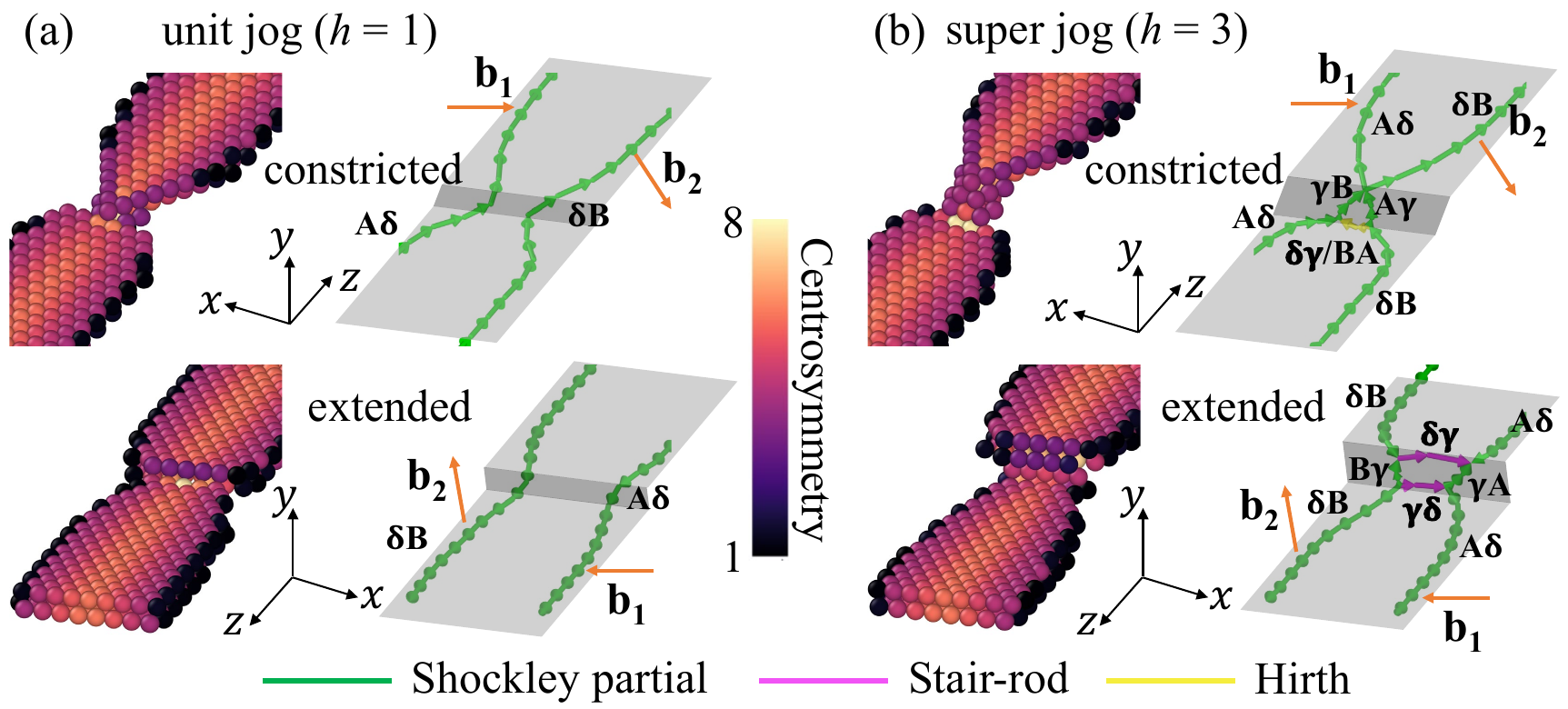}
    \caption{
    Atomic structures and corresponding dislocation line geometries of
    (a) unit jog pair and (b) super-jog pair.
    The dislocation segments on the glide plane dissociate into Shockley partials with Burgers vectors $\mathbf{b}_1$ and $\mathbf{b}_2$ (orange arrows).
    In the jog pair, one jog is extended while the other is constricted.
    The stacking fault atoms are colored according to their Centrosymmetry Parameters (CSP) with 12-neighbor FCC structure.
    Burgers vectors of the partial dislocation lines are shown for super-jog pair.
    }
    \label{fig:jog_pair}
\end{figure}

We obtain relaxed configuration of the jogged edge dislocations through energy minimization, and analyze the atomistic structure of the stacking fault and dislocation jogs using the OVITO~\cite{stukowski2012msmse} software.
\autoref{fig:jog_pair} illustrates the dissociated jog structures of the unit-jog pair configuration ($h=1$) and the super-jog configuration ($h=3$).
For both cases, the jog pair consists of one constricted jog and one extended jog,
consistent with the previous MD simulation of FCC jogged dislocations~\cite{abu-odeh2020acta, rodney2000prb, wang2025scripta}.
We marked the partial dislocations of both the unit and super jog cases with the Thompson's tetrahedral notation~\cite{anderson2017theory}, as shown in \autoref{fig:jog_pair}.
The edge dislocation dissociates on the slip plane of $ABC$ ($\delta$) while the jogs lie on another slip plane $ABD$ ($\gamma$).
Both planes are parallel with the perfect Burgers vector direction $\mathbf{b}$ and thus the jogs on the edge dislocations are glissile.
On the one hand, we note that the dislocation structures for the unit jogs are difficult to be represented by the Thompson's notation because the height of the jogs is only one lattice size.
This one lattice step of the stacking fault is named the `jog line'~\cite{anderson2017theory} structure, from which one can still distinguish constricted and extended jog structures.
%
On the other hand, the superjog configurations can be more easily marked following the Thompson's notation, consistent with the analysis from the author's previous publication~\cite{wang2025scripta}.
According to the dislocation analysis, the constricted jog in the superjog case is dissociated into a triangle-like shape with two Shockley partials and one Hirth partial,
where only the joined node of $\gamma B$ and $A\gamma$ serves as the source of the jog dragging.
Therefore, the dragging effect from the constricted jog may be considered to be independent of the height $h$ of the super jog (as long as $h$ is sufficiently large).
This observation supports the use of the same fitted value of the Peierls energy barrier $H_{\rm 0}^{\rm p}$ for superjogs of $h=3$ and $h=5$, which is higher than the value for the case of unit jog ($h=1$).

\begin{figure}[!ht]
    \centering
    \includegraphics[width=0.5\linewidth]{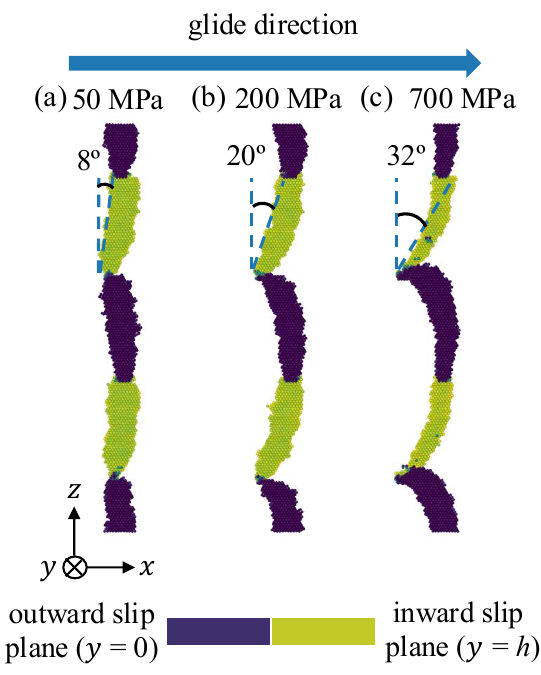}
    \caption{Atomic configurations during the motion of jogged edge dislocation ($L9h3$) at 300 K under the applied stresses of (a) $\SI{50}{MPa}$ (Regime I), (b) $\SI{200}{MPa}$ (Regime II) and (c) $\SI{700}{MPa}$ (Regime III), respectively.}
    \label{fig:visualization}
\end{figure}

\autoref{fig:visualization} illustrates the dislocation structure of $L9h3$ at $\qty{300}{K}$ with different applied stresses.
\autoref{fig:visualization} (a), (b), and (c) correspond to the stress in Regimes I, II, and III, respectively.
It is readily seen that the jog dragging effect comes primarily from the constricted jog, while the extended jogs are moving together with the dislocation and do not show obvious dragging effects.
As a result, the constricted jogs act as pinning points, causing the dislocation arms to bend.
It is interesting to note that the dislocation line bending around the constricted jog is asymmetric.  Instead, there is a tendency to tilt in the $+z$ direction.
%
This asymmetric bending produces a climb force on the constricted jog in the $+z$ direction.
Under high stress conditions (usually above Regime II), the constricted jog is observed to climb accompanied with vacancy emission, which is discussed in our previous publication~\cite{wang2025scripta}.
Jog climb and vacancy emission are not discussed much in this paper, whose focus is on the mobility law of the jogged dislocation in stress Regimes I and II where this phenomenon is rarely observed in our MD simulations.

\subsection{Stress-dependent Peierls barrier for jogged dislocation}
\label{sec:Peierls_stress}

%
\begin{figure}[!ht]
    \centering
    \includegraphics[width=0.99\linewidth]{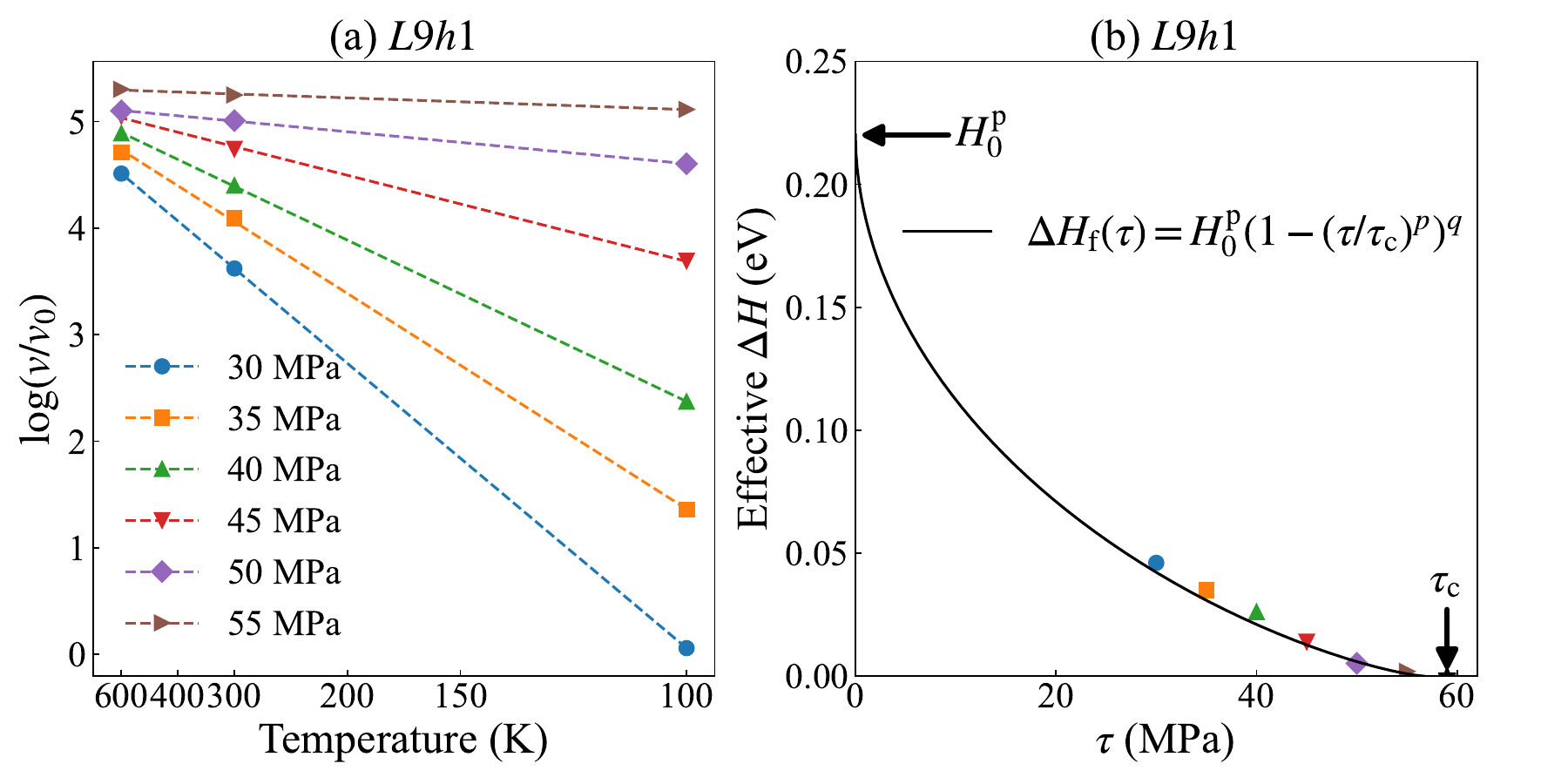}
    \caption{(a) Dislocation velocity of the $L9h1$ configuration at different temperatures calculated under different applied stresses using MD simulations. $v_0 =$ \qty{1}{m/s}.
    The slope of the Arrhenius curve represents the effective activation enthalpy $\Delta H$.
    (b) Effective enthalpy (symbols) as a function of applied stress extracted from (a) together with Eq.~(\ref{eq:DHf}) (solid line).
    }
    \label{fig:thermal_activation}
\end{figure}

Here we take a closer look at the dislocation mobility in the thermally activated regime (Regime I) and provide some justification to the proposed functional form.
\autoref{fig:thermal_activation}(a) is the Arrhenius plot of the dislocation velocity for the unit jog configuration ($L9h1$).
The natural log of the dislocation velocity is plotted as a function of inverse temperature for different applied stresses.
The slope of each dashed line is the effective activation enthalpy $\Delta H$ at the corresponding applied stress $\tau$.
They are plotted as symbols in \autoref{fig:thermal_activation}(b), together with the activation enthalpy $\Delta H_{\rm f}$ as a function of applied stress $\tau$, Eq.~(\ref{eq:DHf}), in a solid line.
We note that $\Delta H_{\rm f}(\tau)$ equals $H_0^{\rm P}$ at $\tau = 0$, and $\Delta H_{\rm f}(\tau)$ reduces to zero at $\tau = \tau_{\rm c}$.
The agreement between the symbols and the solid line in \autoref{fig:thermal_activation}(b) provides support for the proposed functional form of the mobility law, Eqs.~(\ref{eqn:thermally_activated_regime})-(\ref{eqn:DHf_DHb}).

%
%
%
%
%
%
%

\section{Conclusions}
\label{sec:conclusions}
The study of dislocation mobility, particularly in jogged dislocations, is crucial for understanding the mechanical behavior of materials under stress.
Jogged dislocations play a significant role in the plastic deformation mechanisms of crystalline materials.
The capability to accurately model the movement and interactions of these dislocations through DDD simulations offers valuable insights into material strength and ductility. 
In this work, large-scale MD simulations are performed to investigate jog effects on dislocation mobility in single-crystal nickel. Edge dislocations with different jog spacings (or densities) and heights are considered and compared to their counterpart without jogs to demonstrate jog effects.
Such MD simulations are also repeated at various temperatures to uncover temperature effect on jogged dislocation motion. 

For a straight dislocation without jogs, the dislocation velocity initially increases linearly with the applied stress and then gradually approaches a plateau near the sound velocity due to the relativistic effect.
In contrast, a jogged dislocation initially exhibits jerky motion at low-stress (Regime I), due to the thermally activated mechanism of motion. As the stress increases to a medium level (Regime II), the velocity-stress relation transitions to a linear regime characteristic of the phonon drag mechanism.
At higher stress (Regime III), the dislocation velocity-stress relation becomes non-linear again due to relativistic effects and approaches a plateau limited by the sound velocity.
Interestingly, the stress-velocity curves for a given jogged dislocation configuration at different temperatures all intersect at an invariant point $(\tau_{\rm c}, v_{\rm c})$, which also provides a convenient demarcation of Regimes I and II.
We propose a physics-informed mobility expression, based on the invariant point, to describe both the temperature and stress dependence of the jogged dislocation velocity in Regimes I and II.
This simple form of the mobility expression provides a quantitative description of the non-linear mobility law in these regimes.
This mobility law model is important for creating DDD simulations that accurately represent the underlying physics of dislocation evolution and strain hardening in FCC crystals.

\newpage

\textbf{Acknowledgements}

This work was supported by the National Science Foundation under Award Number DMREF 2118522 (W.J. and W.C.).
Y.W. was supported by the Stanford Energy Postdoctoral Fellowship and the Precourt Institute for Energy.

\bibliographystyle{elsarticle-num-names}
\bibliography{ref.bib}

\begin{thebibliography}{28}
\expandafter\ifx\csname natexlab\endcsname\relax\def\natexlab#1{#1}\fi
\providecommand{\url}[1]{\texttt{#1}}
\providecommand{\href}[2]{#2}
\providecommand{\path}[1]{#1}
\providecommand{\DOIprefix}{doi:}
\providecommand{\ArXivprefix}{arXiv:}
\providecommand{\URLprefix}{URL: }
\providecommand{\Pubmedprefix}{pmid:}
\providecommand{\doi}[1]{\href{http://dx.doi.org/#1}{\path{#1}}}
\providecommand{\Pubmed}[1]{\href{pmid:#1}{\path{#1}}}
\providecommand{\bibinfo}[2]{#2}
\ifx\xfnm\relax \def\xfnm[#1]{\unskip,\space#1}\fi
\bibitem[{Hirth and Lothe(1982)}]{hirth1982theory}
\bibinfo{author}{J.~P. Hirth}, \bibinfo{author}{J.~Lothe},
  \bibinfo{title}{Theory of dislocations}, \bibinfo{edition}{2nd} ed.,
  \bibinfo{publisher}{Wiley}, \bibinfo{address}{New York},
  \bibinfo{year}{1982}.
\bibitem[{Hull and Bacon(2011)}]{hull2011introduction}
\bibinfo{author}{D.~Hull}, \bibinfo{author}{D.~J. Bacon},
  \bibinfo{title}{Introduction to dislocations}, \bibinfo{edition}{5th} ed.,
  \bibinfo{publisher}{Butterworth Heinemann, Elsevier},
  \bibinfo{address}{Amsterdam Heidelberg}, \bibinfo{year}{2011}.
\bibitem[{Messerschmidt(1970)}]{messerschmidt1970pssb}
\bibinfo{author}{U.~Messerschmidt},
\newblock \bibinfo{title}{A model of the temperature dependent part of stage i
  work-hardening due to jog-dragging (i)},
\newblock \bibinfo{journal}{Phys. Status Solidi. (b)} \bibinfo{volume}{41}
  (\bibinfo{year}{1970}) \bibinfo{pages}{549--563}.
\bibitem[{Messerschmidt(1971)}]{messerschmidt1971pssb}
\bibinfo{author}{U.~Messerschmidt},
\newblock \bibinfo{title}{A model of the temperature dependent part of stage i
  work-hardening due to jog-dragging (ii)},
\newblock \bibinfo{journal}{Phys. Status Solidi. (b)} \bibinfo{volume}{48}
  (\bibinfo{year}{1971}) \bibinfo{pages}{781--790}.
\bibitem[{Messerschmidt(2010)}]{messerschmidt2010dislocation}
\bibinfo{author}{U.~Messerschmidt}, \bibinfo{title}{Dislocation dynamics during
  plastic deformation}, volume \bibinfo{volume}{129},
  \bibinfo{publisher}{Springer Science \& Business Media},
  \bibinfo{year}{2010}.
\bibitem[{Abu-Odeh et~al.(2020)Abu-Odeh, Cottura, and Asta}]{abu-odeh2020acta}
\bibinfo{author}{A.~Abu-Odeh}, \bibinfo{author}{M.~Cottura},
  \bibinfo{author}{M.~Asta},
\newblock \bibinfo{title}{Insights into dislocation climb efficiency in {FCC}
  metals from atomistic simulations},
\newblock \bibinfo{journal}{Acta Mater.} \bibinfo{volume}{193}
  (\bibinfo{year}{2020}) \bibinfo{pages}{172--181}.
\bibitem[{Rodney and Martin(2000)}]{rodney2000prb}
\bibinfo{author}{D.~Rodney}, \bibinfo{author}{G.~Martin},
\newblock \bibinfo{title}{Dislocation pinning by glissile interstitial loops in
  a nickel crystal: A molecular-dynamics study},
\newblock \bibinfo{journal}{Phys. Rev. B} \bibinfo{volume}{61}
  (\bibinfo{year}{2000}) \bibinfo{pages}{8714--8725}.
\bibitem[{Wang et~al.(2025)Wang, Jian, and Cai}]{wang2025scripta}
\bibinfo{author}{Y.~Wang}, \bibinfo{author}{W.-R. Jian},
  \bibinfo{author}{W.~Cai},
\newblock \bibinfo{title}{Room-temperature vacancy emission from jog on edge
  dislocation in fcc nickel under glide force},
\newblock \bibinfo{journal}{Scr. Mater.} \bibinfo{volume}{260}
  (\bibinfo{year}{2025}) \bibinfo{pages}{116597}.
\bibitem[{Jian et~al.(2022)Jian, Xu, Su, and Beyerlein}]{jian2022acta}
\bibinfo{author}{W.-R. Jian}, \bibinfo{author}{S.~Xu}, \bibinfo{author}{Y.~Su},
  \bibinfo{author}{I.~J. Beyerlein},
\newblock \bibinfo{title}{Energetically favorable dislocation/nanobubble bypass
  mechanism in irradiation conditions},
\newblock \bibinfo{journal}{Acta Mater.} \bibinfo{volume}{230}
  (\bibinfo{year}{2022}) \bibinfo{pages}{117849}.
\bibitem[{Kondo et~al.(2025)Kondo, Shibata, and Ikuhara}]{kondo2025scripta}
\bibinfo{author}{S.~Kondo}, \bibinfo{author}{N.~Shibata},
  \bibinfo{author}{Y.~Ikuhara},
\newblock \bibinfo{title}{Direct observations of jog formation and drag caused
  by screw-screw dislocation interaction},
\newblock \bibinfo{journal}{Scr. Mater.} \bibinfo{volume}{258}
  (\bibinfo{year}{2025}) \bibinfo{pages}{116513}.
\bibitem[{Bulatov et~al.(2006)Bulatov, Bulatov, and Cai}]{bulatov2006computer}
\bibinfo{author}{V.~V. Bulatov}, \bibinfo{author}{V.~Bulatov},
  \bibinfo{author}{W.~Cai}, \bibinfo{title}{Computer simulations of
  dislocations}, volume~\bibinfo{volume}{3}, \bibinfo{publisher}{Oxford
  University Press on Demand}, \bibinfo{year}{2006}.
\bibitem[{Bertin et~al.(2020)Bertin, Sills, and Cai}]{bertin2020armr}
\bibinfo{author}{N.~Bertin}, \bibinfo{author}{R.~B. Sills},
  \bibinfo{author}{W.~Cai},
\newblock \bibinfo{title}{Frontiers in the simulation of dislocations},
\newblock \bibinfo{journal}{Annu. Rev. Mater. Res.} \bibinfo{volume}{50}
  (\bibinfo{year}{2020}) \bibinfo{pages}{437--464}.
\bibitem[{Arsenlis et~al.(2007)Arsenlis, Cai, Tang, Rhee, Oppelstrup, Hommes,
  Pierce, and Bulatov}]{arsenlis2007msmse}
\bibinfo{author}{A.~Arsenlis}, \bibinfo{author}{W.~Cai},
  \bibinfo{author}{M.~Tang}, \bibinfo{author}{M.~Rhee},
  \bibinfo{author}{T.~Oppelstrup}, \bibinfo{author}{G.~Hommes},
  \bibinfo{author}{T.~G. Pierce}, \bibinfo{author}{V.~V. Bulatov},
\newblock \bibinfo{title}{Enabling strain hardening simulations with
  dislocation dynamics},
\newblock \bibinfo{journal}{Modelling Simul. Mater. Sci. Eng.}
  \bibinfo{volume}{15} (\bibinfo{year}{2007}) \bibinfo{pages}{553}.
\bibitem[{Li et~al.(2023)Li, Ghoniem, Baker, {Ramirez Flores}, Black,
  Hollenbeck, and Po}]{li2023acta}
\bibinfo{author}{Y.~Li}, \bibinfo{author}{N.~Ghoniem},
  \bibinfo{author}{K.~Baker}, \bibinfo{author}{B.~{Ramirez Flores}},
  \bibinfo{author}{T.~Black}, \bibinfo{author}{J.~Hollenbeck},
  \bibinfo{author}{G.~Po},
\newblock \bibinfo{title}{A coupled vacancy diffusion-dislocation dynamics
  model for the climb-glide motion of jogged screw dislocations},
\newblock \bibinfo{journal}{Acta Mater.} \bibinfo{volume}{244}
  (\bibinfo{year}{2023}) \bibinfo{pages}{118546}.
\bibitem[{Akhondzadeh et~al.(2023)Akhondzadeh, Kang, Sills, Ramesh, and
  Cai}]{akhondzadeh2023acta}
\bibinfo{author}{S.~Akhondzadeh}, \bibinfo{author}{M.~Kang},
  \bibinfo{author}{R.~B. Sills}, \bibinfo{author}{K.~Ramesh},
  \bibinfo{author}{W.~Cai},
\newblock \bibinfo{title}{Direct comparison between experiments and dislocation
  dynamics simulations of high rate deformation of single crystal copper},
\newblock \bibinfo{journal}{Acta Mater.} \bibinfo{volume}{250}
  (\bibinfo{year}{2023}) \bibinfo{pages}{118851}.
\bibitem[{Cho et~al.(2017)Cho, Molinari, and Anciaux}]{cho2017ijp}
\bibinfo{author}{J.~Cho}, \bibinfo{author}{J.-F. Molinari},
  \bibinfo{author}{G.~Anciaux},
\newblock \bibinfo{title}{Mobility law of dislocations with several character
  angles and temperatures in {FCC} aluminum},
\newblock \bibinfo{journal}{Int. J. Plast.} \bibinfo{volume}{90}
  (\bibinfo{year}{2017}) \bibinfo{pages}{66--75}.
\bibitem[{Dang et~al.(2019)Dang, Bamney, Bootsita, Capolungo, and
  Spearot}]{dang2019acta}
\bibinfo{author}{K.~Dang}, \bibinfo{author}{D.~Bamney},
  \bibinfo{author}{K.~Bootsita}, \bibinfo{author}{L.~Capolungo},
  \bibinfo{author}{D.~E. Spearot},
\newblock \bibinfo{title}{Mobility of dislocations in aluminum: Faceting and
  asymmetry during nanoscale dislocation shear loop expansion},
\newblock \bibinfo{journal}{Acta Mater.} \bibinfo{volume}{168}
  (\bibinfo{year}{2019}) \bibinfo{pages}{426--435}.
\bibitem[{Olmsted et~al.(2005)Olmsted, Hector, Curtin, and
  Clifton}]{Olmsted2005msmse}
\bibinfo{author}{D.~L. Olmsted}, \bibinfo{author}{L.~G. Hector},
  \bibinfo{author}{W.~A. Curtin}, \bibinfo{author}{R.~J. Clifton},
\newblock \bibinfo{title}{Atomistic simulations of dislocation mobility in
  {Al}, {Ni} and {Al}/{Mg} alloys},
\newblock \bibinfo{journal}{Modelling Simul. Mater. Sci. Eng.}
  \bibinfo{volume}{13} (\bibinfo{year}{2005}) \bibinfo{pages}{371}.
\bibitem[{Queyreau et~al.(2011)Queyreau, Marian, Gilbert, and
  Wirth}]{queyreau2011prb}
\bibinfo{author}{S.~Queyreau}, \bibinfo{author}{J.~Marian},
  \bibinfo{author}{M.~R. Gilbert}, \bibinfo{author}{B.~D. Wirth},
\newblock \bibinfo{title}{Edge dislocation mobilities in bcc {Fe} obtained by
  molecular dynamics},
\newblock \bibinfo{journal}{Phys. Rev. B} \bibinfo{volume}{84}
  (\bibinfo{year}{2011}) \bibinfo{pages}{064106}.
\bibitem[{Chang et~al.(2002)Chang, Cai, Bulatov, and
  Yip}]{chang_molecular_2002}
\bibinfo{author}{J.~Chang}, \bibinfo{author}{W.~Cai}, \bibinfo{author}{V.~V.
  Bulatov}, \bibinfo{author}{S.~Yip},
\newblock \bibinfo{title}{Molecular dynamics simulations of motion of edge and
  screw dislocations in a metal},
\newblock \bibinfo{journal}{Comput. Mater. Sci.} \bibinfo{volume}{23}
  (\bibinfo{year}{2002}) \bibinfo{pages}{111--115}.
\bibitem[{Groh et~al.(2009)Groh, Marin, Horstemeyer, and
  Bammann}]{groh2009msmse}
\bibinfo{author}{S.~Groh}, \bibinfo{author}{E.~B. Marin},
  \bibinfo{author}{M.~F. Horstemeyer}, \bibinfo{author}{D.~J. Bammann},
\newblock \bibinfo{title}{Dislocation motion in magnesium: a study by molecular
  statics and molecular dynamics},
\newblock \bibinfo{journal}{Modelling Simul. Mater. Sci. Eng.}
  \bibinfo{volume}{17} (\bibinfo{year}{2009}) \bibinfo{pages}{075009}.
\bibitem[{Thompson et~al.(2022)Thompson, Aktulga, Berger, Bolintineanu, Brown,
  Crozier, in~'t Veld, Kohlmeyer, Moore, Nguyen, Shan, Stevens, Tranchida,
  Trott, and Plimpton}]{thompson2022cpc}
\bibinfo{author}{A.~P. Thompson}, \bibinfo{author}{H.~M. Aktulga},
  \bibinfo{author}{R.~Berger}, \bibinfo{author}{D.~S. Bolintineanu},
  \bibinfo{author}{W.~M. Brown}, \bibinfo{author}{P.~S. Crozier},
  \bibinfo{author}{P.~J. in~'t Veld}, \bibinfo{author}{A.~Kohlmeyer},
  \bibinfo{author}{S.~G. Moore}, \bibinfo{author}{T.~D. Nguyen},
  \bibinfo{author}{R.~Shan}, \bibinfo{author}{M.~J. Stevens},
  \bibinfo{author}{J.~Tranchida}, \bibinfo{author}{C.~Trott},
  \bibinfo{author}{S.~J. Plimpton},
\newblock \bibinfo{title}{{LAMMPS} - a flexible simulation tool for
  particle-based materials modeling at the atomic, meso, and continuum scales},
\newblock \bibinfo{journal}{Comput. Phys. Commun.} \bibinfo{volume}{271}
  (\bibinfo{year}{2022}) \bibinfo{pages}{108171}.
\bibitem[{Angelo et~al.(1995)Angelo, Moody, and Baskes}]{angelo1995msmse}
\bibinfo{author}{J.~E. Angelo}, \bibinfo{author}{N.~R. Moody},
  \bibinfo{author}{M.~I. Baskes},
\newblock \bibinfo{title}{Trapping of hydrogen to lattice defects in nickel},
\newblock \bibinfo{journal}{Model. Simul. Mater. Sci. Eng.} \bibinfo{volume}{3}
  (\bibinfo{year}{1995}) \bibinfo{pages}{289}.
\bibitem[{Kang et~al.(2012)Kang, Bulatov, and Cai}]{kang_singular_2012}
\bibinfo{author}{K.~Kang}, \bibinfo{author}{V.~V. Bulatov},
  \bibinfo{author}{W.~Cai},
\newblock \bibinfo{title}{Singular orientations and faceted motion of
  dislocations in body-centered cubic crystals},
\newblock \bibinfo{journal}{Proc. Natl. Acad. Sci.} \bibinfo{volume}{109}
  (\bibinfo{year}{2012}) \bibinfo{pages}{15174--15178}.
  \bibinfo{note}{Publisher: Proceedings of the National Academy of Sciences}.
\bibitem[{Po et~al.(2016)Po, Cui, Rivera, Cereceda, Swinburne, Marian, and
  Ghoniem}]{po2016acta}
\bibinfo{author}{G.~Po}, \bibinfo{author}{Y.~Cui}, \bibinfo{author}{D.~Rivera},
  \bibinfo{author}{D.~Cereceda}, \bibinfo{author}{T.~D. Swinburne},
  \bibinfo{author}{J.~Marian}, \bibinfo{author}{N.~Ghoniem},
\newblock \bibinfo{title}{A phenomenological dislocation mobility law for bcc
  metals},
\newblock \bibinfo{journal}{Acta Mater.} \bibinfo{volume}{119}
  (\bibinfo{year}{2016}) \bibinfo{pages}{123--135}.
\bibitem[{Mecking and Kocks(1981)}]{mecking1981acta}
\bibinfo{author}{H.~Mecking}, \bibinfo{author}{U.~F. Kocks},
\newblock \bibinfo{title}{Kinetics of flow and strain-hardening},
\newblock \bibinfo{journal}{Acta Metall.} \bibinfo{volume}{29}
  (\bibinfo{year}{1981}) \bibinfo{pages}{1865--1875}.
\bibitem[{Stukowski et~al.(2012)Stukowski, Bulatov, and
  Arsenlis}]{stukowski2012msmse}
\bibinfo{author}{A.~Stukowski}, \bibinfo{author}{V.~V. Bulatov},
  \bibinfo{author}{A.~Arsenlis},
\newblock \bibinfo{title}{Automated identification and indexing of dislocations
  in crystal interfaces},
\newblock \bibinfo{journal}{Modelling Simul. Mater. Sci. Eng.}
  \bibinfo{volume}{20} (\bibinfo{year}{2012}) \bibinfo{pages}{085007}.
\bibitem[{Anderson et~al.(2017)Anderson, Hirth, and Lothe}]{anderson2017theory}
\bibinfo{author}{P.~M. Anderson}, \bibinfo{author}{J.~P. Hirth},
  \bibinfo{author}{J.~Lothe}, \bibinfo{title}{Theory of Dislocations},
  \bibinfo{publisher}{Cambridge University Press}, \bibinfo{year}{2017}.

\end{thebibliography}

\end{document}